\documentclass{article}
\usepackage[T1]{fontenc}
\usepackage[utf8]{inputenc}
\usepackage{ismir} 
\usepackage{amsmath,cite}
\usepackage{amssymb}      
\usepackage{graphicx}
\usepackage{color}
\usepackage[hyphens]{url}

\usepackage{subcaption}
\usepackage{tabularray}
\usepackage{multirow}
\usepackage{booktabs}
\usepackage{makecell}
\usepackage{siunitx}
\usepackage{dashrule}
\usepackage{xspace}

\title{Emergent musical properties of a transformer under contrastive self-supervised learning}

\multauthor
{Yuexuan Kong$^{1, 2}$ \hspace{1cm} Gabriel Meseguer-Brocal$^1$ \hspace{1cm} Vincent Lostanlen$^2$} { \bfseries{Mathieu Lagrange$^2$ \hspace{1cm} Romain Hennequin$^1$}\\
 $^1$ Deezer Research, Paris, France\\
$^2$ Nantes Université, École Centrale Nantes, CNRS, LS2N, UMR 6004, F-44000 Nantes, France\\
{\tt\small ykong@deezer.com}
}

\newcommand{\melt}{ViT-1D\xspace}

\def\authorname{Y. Kong et al.}
\begin{document}
\captionsetup{font=small}
\maketitle

\begin{abstract}

In music information retrieval (MIR), contrastive self-supervised learning for general-purpose representation models is effective for global tasks such as automatic tagging.
However, for local tasks such as chord estimation, it is widely assumed that contrastively trained general-purpose self-supervised models are inadequate and that more sophisticated SSL is necessary; e.g., masked modeling.
Our paper challenges this assumption by revealing the potential of contrastive SSL paired with a transformer in local MIR tasks.
We consider a lightweight vision transformer with one-dimensional patches in the time--frequency domain (\melt) and train it with simple contrastive SSL through normalized temperature-scaled cross-entropy loss (NT-Xent).
Although NT-Xent operates only over the class token, we observe that, potentially thanks to weight sharing, informative musical properties emerge in ViT-1D's sequence tokens.
On global tasks, the temporal average of class and sequence tokens offers a performance increase compared to the class token alone, showing useful properties in the sequence tokens.
On local tasks, sequence tokens perform unexpectedly well, despite not being specifically trained for.
Furthermore, high-level musical features such as onsets emerge from layer-wise attention maps and self-similarity matrices show different layers capture different musical dimensions.
Our paper does not focus on improving performance but advances the musical interpretation of transformers and sheds light on some overlooked abilities of contrastive SSL paired with transformers for sequence modeling in MIR.

\end{abstract}

\section{Introduction}
\label{sec:introduction}

\begin{figure*}
    \centering
    \includegraphics[width=0.875\linewidth]{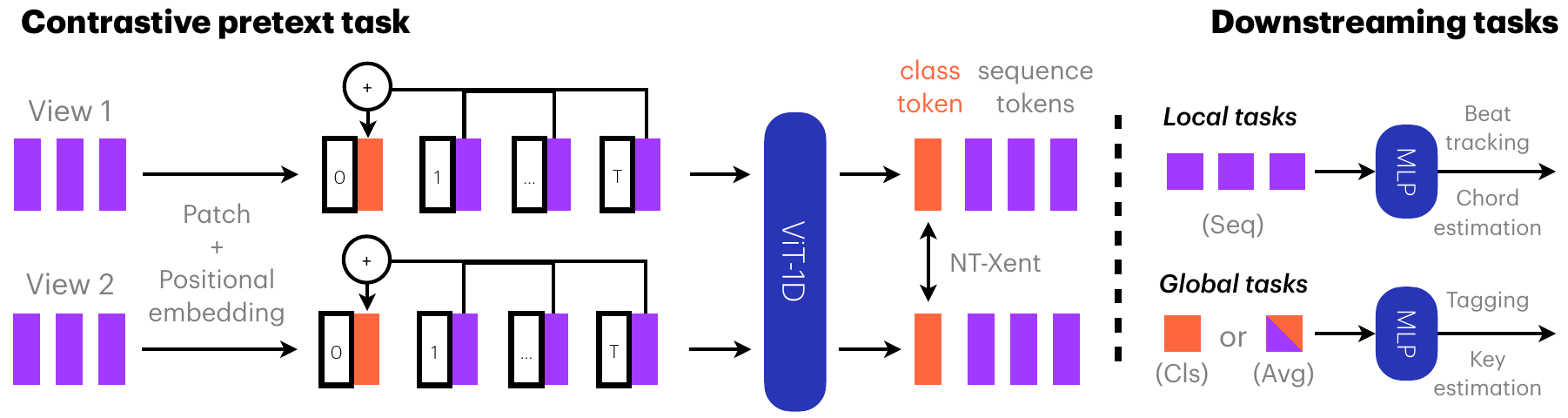}
    \caption{Contrastive pre-training and probing for downstream tasks. The inputs (left) are mel-spectrograms, patched along vertical slices of frequency bins per time frame. Positional encoding and a class token (learnable parameters with the average of the sequence tokens) are added. NT-Xent loss is applied only to the class token. For downstream tasks, sequence tokens \texttt{(Seq)} (excluding the class token) are used for local tasks, while the class token \texttt{(Cls)} or the average of all tokens \texttt{(Avg)} is used for global tasks.} 
    \label{fig:framework}
\end{figure*}

We may categorize tasks in music information retrieval (MIR) as either local or global.
Global tasks, such as music tagging and key estimation, are time-shift invariant and require a single prediction per piece of music.
Local tasks, such as beat tracking and chord estimation, are time-shift equivariant and require frame-wise predictions, with a frame rate typically higher than \SI{1}{\hertz} \cite{lostanlen2017convolutional}.

To address these tasks, self-supervised learning (SSL) has recently emerged as a powerful alternative to supervised learning in MIR. SSL enables a model to learn informative representations through a pretext task without requiring labeled data. While these pretext tasks may not have direct practical relevance, solving them requires the model to capture one or various musical dimensions \cite{PESTO, quinton2022equivariant,kong2024stone, kongskey, meseguer2020data, ma2024foundation}. In general-purpose models, these learned representations are then useful for many different downstream tasks, requiring only a small amount of supervision. 

In general-purpose SSL for MIR, CLMR \cite{spijkervet2021contrastive} and MULE \cite{mccallum2022supervised} marked a first step forward, following the adoption of contrastive pretext task in computer vision \cite{simclr, chen2021exploring}. In contrastive learning, the model is enforced by a loss to project positive pair samples close together in the embedding space and pushing negative samples far apart. Their results showed the potential of contrastive SSL to generalize across various global music tasks. However, due to the properties of convolutional neural networks and global pooling layers, both models capture global music representations that summarize the entire sequence rather than preserving information at each time step.
More general-purpose SSL research further developed on contrastive pretext tasks by using a momentum-based paradigm \cite{zhao2022s3t}, combining different musical stems \cite{garoufis2023multisource}, analyzing transformation in embedding space \cite{mccallum2024effect}, and developing more effective training strategies \cite{choi2022towards}.
Aforementioned papers only evaluate their systems on global tasks.
In contrast, the potential of general-purpose contrastive SSL on local tasks remains understudied.

Generative modeling and masked modeling are widely used for general-purpose SSL at the frame level.
Generative models such as Jukebox \cite{dhariwal2020jukebox} and Music2Latent \cite{pasini2024music2latent} have showed that various musical dimensions are captured in their embedding space used for generation, by evaluating on multiple MIR tasks. 
MERT \cite{li2023mert} is a music representation learning model that resembles the masking training scheme of Hu-BERT \cite{hsu2021hubert} from speech. M2D \cite{niizumi2024m2dx} employs a joint-embedding predictive architecture (JEPA) that jointly predicts from both a masked sample and the original sample. MusicFM \cite{won2024foundation} conducts a comparative study of different masked modeling approaches. 
Although these models handle both global and local tasks well, they have two shortcomings.
First, large-scale architectures are necessary: the number of parameters typically ranges from 58M (Music2latent) to 5B (Jukebox).
Secondly, training these models depends on sophisticated techniques such as exponential moving averages, teacher--student distillation, and multiple loss functions; requiring careful fine-tuning of hyperparameters and large computational resources.

Transformers have been applied to contrastive pretext tasks \cite{wang2022towards, koutini2022efficient} using the AST architecture \cite{gong21b_interspeech} and for multimodal audio-text learning \cite{manco2022contrastive, huang2022mulan}.
In these cases, contrastive loss is applied only to the class token; i.e., a learnable token attached at the beginning of the sequence. Optimization of the loss brings paired audio-audio or audio-text closely in the embedding space.
Computer vision researchers have reported emergent properties when training Vision Transformers (ViTs) \cite{dosovitskiy2021an}. Crucially, such properties do not emerge through supervised pretraining\cite{caron2021emerging}. This approach has proven valuable, not only for global tasks such as classification, but also for local tasks such as image segmentation\cite{caron2021emerging, oquab2023dinov2}.
Attention maps are also studied to show the emergent local properties, providing insights into the local patterns and features which are learned during training. 
However, to our knowledge, these emergent properties in transformer tokens have not yet been explored on local tasks for music.
 

In general-purpose SSL, we notice a gap between contrastive SSL and masked modeling in MIR, particularly regarding the ability of contrastive pretext task to capture both global and local properties. 
This gap leads us to the following questions: have we moved on too quickly from contrastive SSL to more complex approaches? Does it still hold more untapped potential while paired with a transformer? To answer them, we proceed in following ways:
\begin{description}
\item[Pretext task.]{We use a lightweight ViT with 1-D spectrogram patches as token inputs (ViT-1D). We train ViT-1D with with a normalized temperature-scaled cross-entropy loss (NT-Xent) only to the class token of positive and negative pairs (Section \ref{sec:method}).}
\item[Downstream tasks.]{We evaluate the effectiveness of both the class token and sequence tokens on local and global downstream tasks. While the class token is time-invariant due to the pretext task formulation, we show that sequence tokens capture local musical properties (Section \ref{sec:downstream}).}
\item[Emergent properties.]{To understand how local properties are captured, we conduct qualitative and quantitative analyses of attention maps (Section \ref{sec:attention-maps}) and self-similarity matrices (Section \ref{sec:ssm}).}\footnote{Code, checkpoint and more examples for Section \ref{sec:attention-maps} and \ref{sec:ssm} can be found at \url{https://github.com/deezer/emergent-musical-properties-transformer/tree/main.}}
\end{description}

\begin{table*}[]
    \centering
    \small
    \caption{Downstream performance comparison of different models on two global  tasks (music tagging and key estimation) and two local tasks (beat detection and chord estimation).  The row \texttt{Cls} corresponds to probing with the class token while the row \texttt{Avg} refers to probing with the average of all tokens. \texttt{Seq} refers to probing only with sequence tokens. }
    \label{tab:results1}
    \begin{tabular}{c c c ccc c cc}
        \toprule
        & & & \multicolumn{4}{c}{\textsc{Global}} & \multicolumn{2}{c}{\textsc{Local}} \\ 
        \midrule
        & \multirow{2}{*}{\textsc{\#param}}
        & \multirow{2}{*}{\textsc{dim}} 
        & \multicolumn{3}{c}{\textsc{Music tagging}}  
        & \textsc{Key estimation} 
        & \textsc{Beat tracking} 
        & \textsc{Chord estimation} \\  
        
        & & & & {\textsc{map}} & \textsc{roc} & \textsc{w. acc} & \textsc{F-score} & \textsc{acc} \\ 
        \midrule \midrule
    
        \multirow{2}{*}{\textsc{vit-1d}} 
        & \multirow{2}{*}{5.3M} 
        & \multirow{2}{*}{192} 
        & \rule{0pt}{7pt} \texttt{(Cls)} & 0.400 & 0.888 & 0.509 
        & \multirow{2}{*}{\texttt{(Seq)} 0.723} & \multirow{2}{*}{0.319} \\ 
        
        & & \rule{0pt}{10pt} &\texttt{(Avg)} & 0.417 & 0.896 & 0.622 & & \\ 
        \hline
    
        \textsc{CLMR-like\cite{meseguer2024experimental}} & 2.8M 
        & 1024
        & & 0.427 & 0.898 & 0.459 & 0.313 & 0.148 \\ 
        \midrule \midrule
    
        \textsc{m2d}\cite{niizumi2024m2dx} & 89M
        & 3840
        & & {0.479} & 0.918 & 0.531 & 0.794 & 0.322 \\ 
        
        \bottomrule
    \end{tabular}
\end{table*}

\section{Contrastive pretext task}
\label{sec:method}

\textbf{Patching details: }We compute the mel-frequency spectrogram for a segment of duration equal to $d=4$ seconds, obtaining matrices $\boldsymbol{x}$, with 128 frequency bins and a frame rate of $\xi=$ \SI{31.5}{\hertz}. Unlike standard ViT, which uses 2D patches, we extract 1D patches by taking all 128 mel bins from a single frame and apply one convolutional layer $f_{\boldsymbol{p}}$, projecting into an embedding of size $(H_p, W_p) = (192, 1)$ for each patch $\boldsymbol{x}_{p} = f_{\boldsymbol{p}}(\boldsymbol{x})$ and $\boldsymbol{x}_{p} \in \mathbb{R}^{{H_p}\times{W_p}}$. By using 1D patches, each patch is directly connected to all the frequency bins in a time frame. We obtain the patch sequence $\boldsymbol{x}_p$ as $[\boldsymbol{x}_p^{1}, \boldsymbol{x}_p^{2}, ..., \boldsymbol{x}_p^{T}]$ with $T=df=126$, where each patch corresponds to one time frame in the mel-spectrogram. This sequence input to a transformer is commonly named as sequence tokens\cite{dosovitskiy2021an}.

\textbf{Encoder architecture:} We use the original ViT implementation of the smallest version as encoder (with our patching method) with the embedding dimension equals to 192, 12 transformer blocks and 3 attention heads. 
Unlike commonly done in SSL, we attach no disposable projection head to the transformer encoder, which possibly reduces overall performance of the model as adding them during the pretext training benefits downstream tasks\cite{simclr}, in order to focus on the emergent properties purely in the transformer. 
We denote $f_{\boldsymbol{e}}$ for our encoder. 
We prepend a class token, composed by learnable parameters and the average of other tokens, at the beginning of $\boldsymbol{x}_p$. Then, a 2D sinusoidal positional encoding on the frequency and time dimensions is added to all patches including the class token, obtaining the final input tokens of $f_{\boldsymbol{e}}$ as $[\boldsymbol{z}_0^{0}, \boldsymbol{z}_0^{1}, ..., \boldsymbol{z}_0^{T}]$ with $T =126$. We define the output token sequence of transformer block $k$ at time $t$ as $\boldsymbol{z}_k^{t}$ where $0 < k \leq 12$ and $0 \leq t \leq T$. Notably, the class token ($t=0$) is processed in the same manner as sequence tokens: it shares weights with them in the multi-layer perception layers following the attention block. This design allows the class token to integrate and summarize information \cite{dosovitskiy2021an}. The output of the model is $[\boldsymbol{z}_L^{0}, \boldsymbol{z}_L^{1}, ..., \boldsymbol{z}_L^{T}]$ where $\boldsymbol{z}_L^{0}$ is the class token and $L=12$.

\textbf{Normalized temperature-scaled cross entropy loss (NT-Xent).}
For each piece of music, we extract two disjoint segments A and B of four seconds each as a pair of positive samples. No data augmentation is applied.
All other segments from the same batch are negative samples. 
To study the emerging properties in the sequence tokens, we apply the loss only on the class tokens\cite{wang2022towards, huang2022mulan, manco2022contrastive}, defining our loss function for each pair as:
\begin{equation}
    \mathcal{L}_{A, B}(f_{\boldsymbol{e}}) = - \log \frac{\exp(\text{sim}(\boldsymbol{z}_{L, A}^0, \boldsymbol{z}_{L, B}^0) / \tau)}{
    \sum_{k\neq B} \exp(\text{sim}(\boldsymbol{z}_{L, A}^0, \boldsymbol{z}_{L, k}^0) / \tau)}
\end{equation}
where $\boldsymbol{z}_{L, A}^0$ and $\boldsymbol{z}_{L, B}^0$ are the class tokens of the positive pair of segments A and B, $\mathrm{sim}$ is the cosine similarity function, and $\tau=0.1$ is the temperature parameter.

\textbf{Pre-training details}
We pretrain on a subset of Deezer's catalog of music, with a batch size of 256 pairs of 4-second segments, a base learning rate of $3 \times 10^{-4}$ with a cosine decay until $5 \times 10^{-7}$, and train for 300 epochs.

\section{Downstream tasks}
\label{sec:downstream}
We focus on two types of downstream tasks, commonly used in general-purpose SSL for MIR. We select music tagging and key estimation as representative global tasks and we choose beat tracking and chord estimation as examples of local tasks. A good performance on these four tasks requires the model to encode both harmonic and rhythmic representations, and high-level musical concept, on both local an global levels.

As discussed in Section \ref{sec:introduction}, to the best of our knowledge, no prior study has explored the emerging properties of sequence tokens in a transformer in local tasks trained with a contrastive learning framework. As a matter of fact, testing on local tasks may seem counterintuitive, since positive samples are simply two segments from the same piece of music, without any explicit alignment of beats or chords, therefore the class tokens are trained to be time-shift invariant. This raises the possibility that local musical information may not be expected in the token sequence since there are no constraints to encourage this. However, our results challenge this assumption, showing that meaningful local representations do emerge despite the lack of direct supervision at the frame level.

\subsection{Music tagging}

\noindent
\textbf{Datasets.} We use MagnaTagaTune \cite{law2009magnatagatune} with the split proposed by Lee et al.\cite{lee2017sample}.

\noindent
\textbf{Training methods. }We compare two different probing methods that both use a single linear layer:  
\textbf{1)} Probe only on the class token $\boldsymbol{z}_L^{0}$ (referred as \texttt{Cls} in Figure \ref{fig:framework} and in Table \ref{tab:results1});
\textbf{2)} Probe on the average of the entire token sequence $[\boldsymbol{z}_L^{0}, \boldsymbol{z}_L^{1}, ..., \boldsymbol{z}_L^{T}]$, including the class token (referred as \texttt{Avg} in Figure \ref{fig:framework} and in Table \ref{tab:results1}).

\noindent
\textbf{Metrics. }We use the area under the receiver operating characteristic curve (ROC-AUC) and mean average precision (mAP) in their macro-aggregated versions.

\subsection{Music key estimation}
\noindent
\textbf{Datasets.} We use FMAKv2 \cite{kong2024stone} with a 9:1 split between training and validation.
FMAKv2, a derivative of FMAK \cite{wong2023fmak}, contains 5489 songs from the Free Music Archive \cite{defferrard2017fma}, spanning multiple genres.
We test on GiantSteps\cite{knees2022giansteps}, a dataset of 604 electronic dance music tracks.

\noindent
\textbf{Training methods. }We compare the same two training methods (\texttt{Avg} and \texttt{Cls}) as in music tagging.

\noindent
\textbf{Metrics. }We use the weighted accuracy from mir\_eval\cite{raffel2014mir_eval}, which assigns weights to some key prediction errors.

\subsection{Beat tracking}
\noindent
\textbf{Datasets. }We use the Ballroom dataset \cite{gouyon2004ballroom} with a 9:1 split for training and validation, which contains 698 ballroom songs. For testing, we use GTZAN Rhythm \cite{marchand2015gtzanrhythm}, which includes beat annotations for 998 songs across 10 genres.

\noindent
\textbf{Training methods. }We exclude the class token and use only the sequence tokens $[\boldsymbol{z}_L^{1}, ..., \boldsymbol{z}_L^{T}]$, referred as \texttt{Seq} in Figure \ref{fig:framework} and Table \ref{tab:results1}. Since beat tracking typically requires a higher frame rate than \SI{31.5}{\hertz}, we attach two independent heads to each $\boldsymbol{z}_L$, doubling the frame rate to \SI{63}{\hertz}. Additionally, we apply a standard smoothing method for beat tracking, where we increase the values of the two neighboring frames to 0.5 instead of 0.

\noindent
\textbf{Metrics. }We apply a Dynamic Bayesian Network (DBN) for post-processing to obtain beat locations\cite{krebs2015efficient}. We use $F$-score with a tolerance window of \SI{70}{\milli\second} as evaluation metrics from the \textrm{mir\_eval} package \cite{raffel2014mir_eval}.

\subsection{Chord estimation}
\noindent
\textbf{Datasets}. We collect 124 songs from the Real World Computing Pop (RWC-POP) and Schubert Winterreise Dataset (SWD) \cite{weiss2021schubert, goto2002rwc}, limited to one performance per song.
We apply a 8:1:1 split between training, validation, and test. We consider 24 classes of major and minor chords, exclude those that cannot be mapped to these classes (e.g., suspended chords), and include a "no chord" class, resulting in 25 classes. The chord vocabulary is the same as used in MusicFM\cite{won2024foundation}.

\noindent
\textbf{Training methods. }We exclude the class token and use only the sequence tokens $[\boldsymbol{z}_L^{1}, \ldots, \boldsymbol{z}_L^{T}]$. The frame rate of the encoder is sufficient for chord estimation, therefore only one single linear layer is used.

\noindent
\textbf{Metrics. }We use frame-level accuracy over 25 classes.

\section{Results on downstream tasks}
Using the frozen output of the pretrained \melt as input to a trainable linear layer for each task, we study whether sequence tokens capture local properties, despite the class token's time-invariance. We also assess their contribution to global tasks. We compare this to two reference models, pretrained with contrastive learning and masked modeling, evaluating them on the same downstream tasks. It is important to note that during pretext training, our model does not include a projection head after the backbone, which is a common technique used to boost performance on downstream tasks \cite{spijkervet2021contrastive, mccallum2022supervised, meseguer2024experimental}. However, we omit it in order to study more directly the emergent properties of the transformer backbone.


CLMR\cite{spijkervet2021contrastive} is a general-purpose contrastive framework introduced for musical representation learning.  
However, CLMR is trained using pitch shift as data augmentation for positive samples, making it invariant to pitch shift, resulting in low performance in tonality-related tasks. 
Therefore, for a more fair comparison, we use a CLMR-like contrastively trained ResNet \cite{meseguer2024experimental}, without any data augmentation, trained on the same dataset as \melt, with the same sampling rate and audio length.
For local tasks, we upsample its resolution from \SI{0.25}{\hertz} to the resolution of ViT-1D for both local tasks by attaching the necessary number of linear layers. It is important to note that this upsampling process results in much more parameters for downstream training than \melt, which only uses 1 and 2 linear layers respectively.

M2D \cite{niizumi2024m2dx} employs a JEPA architecture, which combines masked modeling with a teacher--student framework trained on general audio. It is the masked modeling model that has the least amount of parameters in Section \ref{sec:introduction}. We use the results as a reference. It produces frame-wise predictions at a rate of \SI{6.3}{\hertz}. To adapt it for beat tracking and chord estimation, we train multiple independent linear layers to upsample to the same frame rate as \melt.

We observe two key findings from Table \ref{tab:results1}. \textbf{1)} Sequence tokens show better performance than CLMR-like model and  comparable results to M2D on local tasks. This suggests the emergence of local and temporal musical representations, in contrast to the time-invariant nature of the class token.
The music tagging performance lags behind CLMR-like and M2D, however both models have a much larger size of embedding dimension, and CLMR-like model uses a projection head after the backbone.
\textbf{2)} For global tasks, performance improves when averaging the class token and all sequence tokens together. This implies that the information encoded in sequence tokens is not entirely captured by the class token alone and that incorporating sequence tokens contributes positively to global tasks.

Local musical properties in the sequence tokens improve performance on global downstream tasks and yield unexpected good results for local tasks.
This shows that, despite \melt being trained with NT-Xent loss only on the class token and the positive sampling strategy making it time-invariant, useful local musical properties still emerge in the sequence tokens. This raises interest in further analyzing the emergent properties across different transformer layers through attention maps (Section \ref{sec:attention-maps}) and in self-similarity matrices (Section \ref{sec:ssm}).

\begin{figure}[t]
    \centering
    \includegraphics[width=\linewidth]{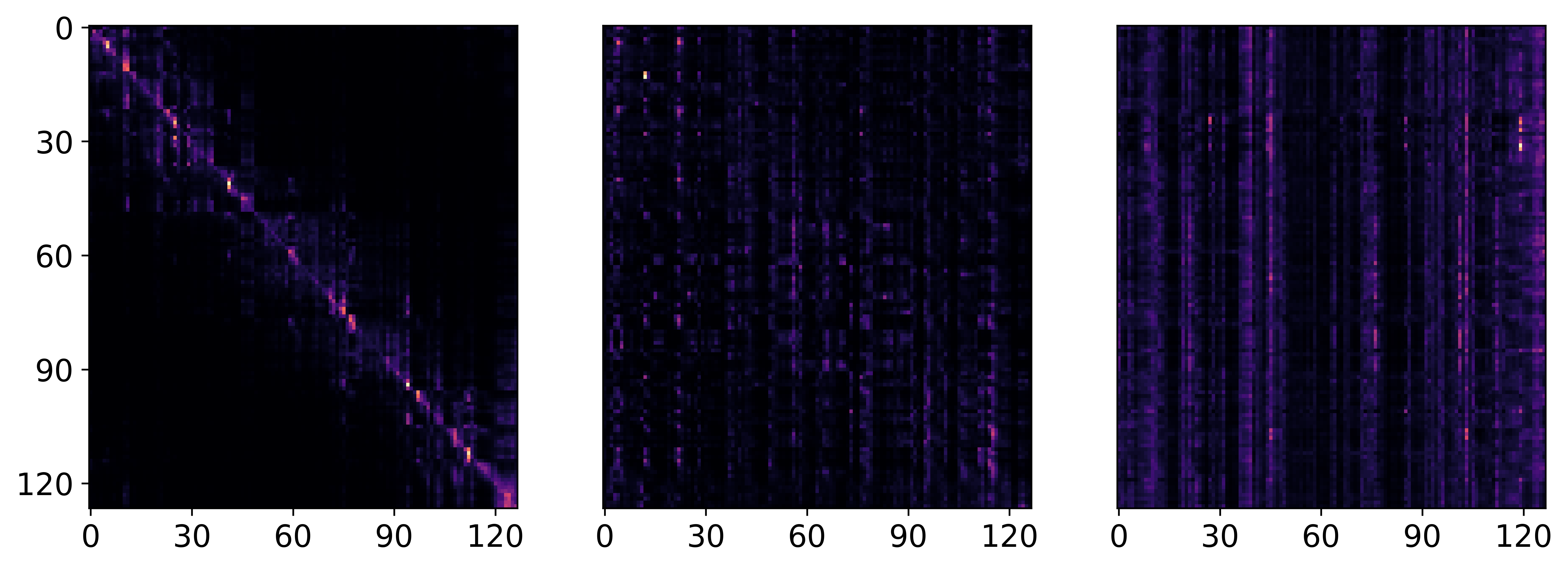}
    \caption{Attention matrices from the 3rd, 9th, and 12th transformer blocks (left to right). Lighter colors at position $[i, j]$ indicate more attention from token $i$ to token $j$. Diagonal lines in the left figure show local attention, while vertical lines across the map in the right figure indicate a shift to global attention in deeper layers.}
    \label{fig:att-layers}
\end{figure}

\section{Properties in attention maps}
\label{sec:attention-maps}
We study the emergent properties of tokens in the transformer across different layers. ViT-1D has 12 layers in total. We select the 3rd, 6th, 9th, and 12th layers as representative points, as they are evenly spaced from shallower to deeper layers. A more comprehensive analysis of all 12 layers, as well as the potential performance gains from leveraging all layers, is left for future work.

\subsection{Qualitative analysis of attention maps}

The attention mechanism directs attention to meaningful token positions during training and are calculated via scaled dot-product self-attention \cite{vaswani2017attention}:
\begin{equation}
    \mathbf{M}_{k}^{h}(\mathbf{Q}_{k}^{h}, \mathbf{K}_{k}^{h}) = \text{softmax} \left( \frac{\mathbf{Q}_{k}^{h}{\mathbf{K}_{k}^{h}}^{\top}}{\sqrt{d}} \right)
\end{equation}
where $0 < k \leq L =12$ is the depth of transformer block, $0 < h \leq 3$ is the index of head and $d=64$ is the dimension of the embeddings of each head. This results in an attention map $\mathbf{M}_{k}^{h} \in \mathbb{R}^{(T+1)\times(T+1)}$ of head $k$ at $h^\mathrm{th}$ transformer block where $T = 126$. In this section, we aim to explore the question: with a simple contrastive pretext task applied to the class token, can attention be guided toward musically meaningful positions in the sequence?

We show the attention matrices from 3rd, 9th and 12th layers of a 4-second polyphonic sample from RWC-Pop in Figure \ref{fig:att-layers}. Although we also study the 6th layer, it is omitted from the figure due to space limit. Similar properties are common across other samples. The value of $\mathbf{M}_{k}^{h}$ at row $i$ and column $j$ presents the attention of token $z_k^i$ on $z_k^j$.

We observe that in the shallow layers of the attention maps, as seen by the presence of short vertical lines along the diagonals, attention is primarily distributed to neighbor tokens. For a given token, only nearby tokens receive attention. However, in deeper layers, vertical lines extend across the entire attention map, indicating that attention is distributed more uniformly across all tokens. For a given token, tokens across the whole sequence can receive attention. A transition from local to global attention is observed from shallower to deeper layers, observed as well in transformers trained for sentence embeddings\cite{9140343}.

\subsection{Alignment of attention maps with onset events}

To quantitatively assess the emergence of temporal properties in attention maps, we use an attention head of an attention block to infer onset event timestamps. We use the MUS subset of the Midi-Aligned Piano Sounds (MAPS-MUS) dataset \cite{emiya2009multipitch}, which contains 30 polyphonic classical piano recordings with aligned MIDI annotations.
We select this dataset and task because it is simpler to build precise hypotheses and interpret attention maps when a single instrument is present and when time-aligned symbolic information is available. However it remains polyphonic, ensuring that the task is still non-trivial.

\begin{figure}[t]
    \centering
    \includegraphics[width=\linewidth]{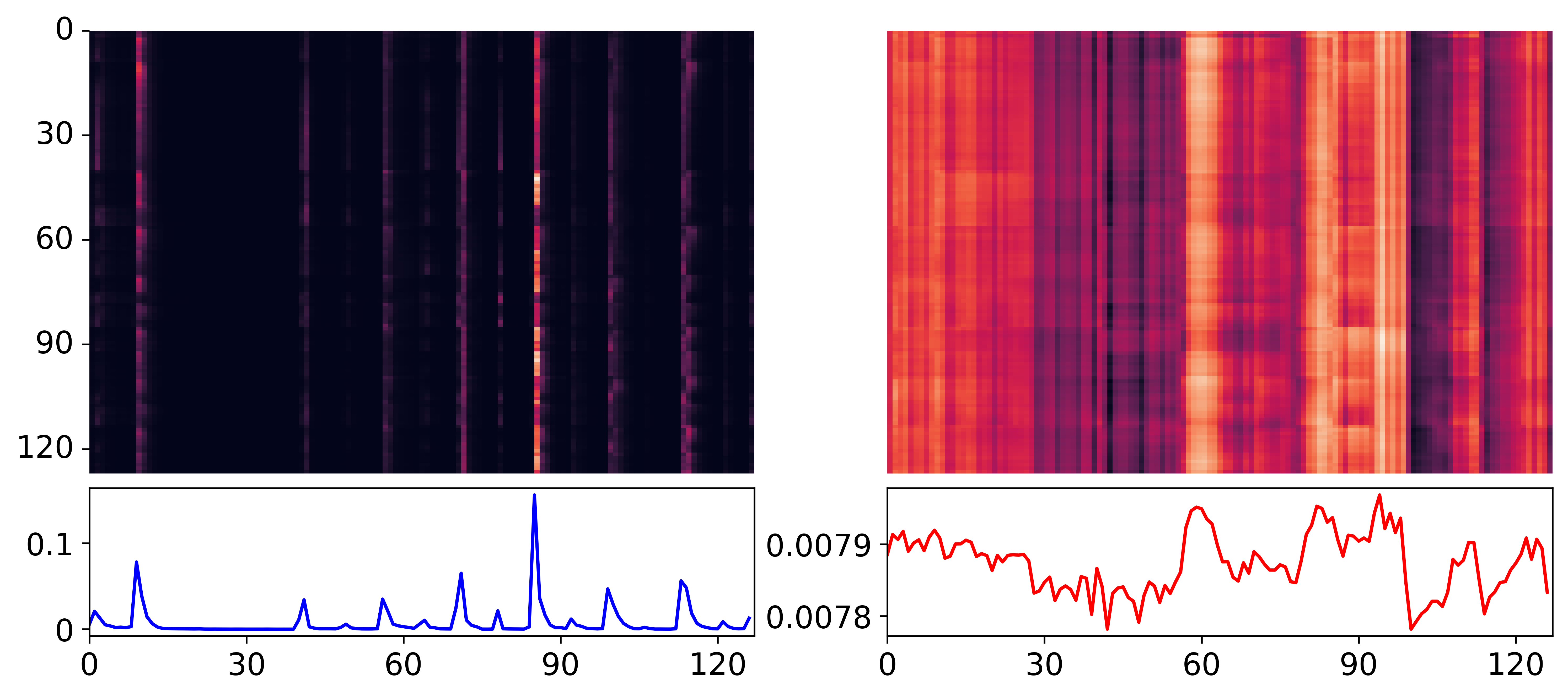}
    \caption{Attention matrices of a trained \melt (top left) and a randomly initialized one (top right), with brighter color indicating higher attention. The scales differ between the top figures, as shown in the bottom plots, which display averaged attention matrices by column. The bottom left shows clear peaks, while the bottom right has similar maximum and minimum values, indicating that attention is evenly distributed at random initialization and becomes more focused on temporal positions during training.}
    \label{fig:act}
\end{figure}

Among the 4 layers we study, similar properties are shown across many heads from 9th and 12th layers, also across multiple models initialized differently. We choose the attention matrix of an attention head from the $9^{\mathrm{th}}$ layer, referred as $\mathbf{M}_{i,j}$ in the following. We exclude the class token, average the attention map per column, and obtain a pseudo-activation function $a(i) = \frac{1}{126} \sum_{j=1}^{126} M_{i,j}$, where $0 < i \leq 126$. This approach is motivated by the observation of vertical lines in deeper layers, indicating that tokens receiving higher attention are similar for all tokens. Figure \ref{fig:act} shows $\mathbf{M}_{i,j}$ and $a(i)$ (left) and a \melt at random initialization (right) for a specific sample. The attention map of the random initialized model exhibits a very narrow value range, and the activation function is almost flat, indicating no meaningful attention was placed at the beginning of the training.

We use the peak picking function from SciPy \cite{2020SciPy-NMeth} on $a(i)$ to obtain onset position and the $F$-score in mir\_eval for evaluation, with a tolerance window of \SI{70}{\milli\second}. For the sake of comparison, we compare this method with spectral flux implementation in librosa and with a \melt at random initialization. We use $F$-score as a metrics with a tolerance window of \SI{70}{\milli\second}, as implemented in \textrm{mir\_eval}\cite{raffel2014mir_eval}.

\begin{table}[t!]
\centering
\resizebox{\columnwidth}{!}{ 
\begin{tabular}{c c c c}
\toprule
\textbf{} & \textsc{Att. map} & \textsc{Random} & \textsc{Spectral flux} \\ \midrule
\textsc{$F$-score} & 0.877 & 0.501 & 0.720 \\ 
\bottomrule
\end{tabular}
}
\caption{$F$-score of onset detection (MAPS-MUS dataset) after peak picking from an attention map of trained \melt (left), compared with a ViT-1D at random initialization (center) and a feature engineering baseline (right).}
\label{tab:fscore}
\end{table}

The comparison between the attention map and the spectral flux method shows a strong alignment between attention and onset events. Furthermore, the temporal properties useful for onset event detection do not appear at random initialization; rather, they emerge during training. A contrastive pretext task applied to the class token alone directs the attention from random to musically relevant positions, without the need for specific training to do so.

\section{Properties in self-similarity matrics of tokens}
\label{sec:ssm}
\subsection{Qualitative analysis}
\label{sec:qual-ssm}

We extract intermediate tokens $[\boldsymbol{z}_k^{1}, \ldots, \boldsymbol{z}_k^{T}]$ at layers $k = 3, 6, 9, 12$ (same as Section \ref{sec:attention-maps}, denoted $\boldsymbol{z}_3$ to $\boldsymbol{z}_{12}$), along with tokens from a randomly initialized \melt{} model, denoted $\boldsymbol{z}_r$. For each $\boldsymbol{z}_k$, we compute a self-similarity matrix (SSM) $\boldsymbol{S}_k[i, j] = \mathrm{sim}(\boldsymbol{z}_k[i], \boldsymbol{z}_k[j])$ using cosine similarity. Due to space limit, SSMs of the 6th and 9th layers are omitted but are available in the GitHub repository.

\begin{figure}[t!]
    \centering
    \resizebox{\columnwidth}{!}{ 
    \begin{subfigure}{0.5\textwidth}
        \centering
        \includegraphics[width=\textwidth]{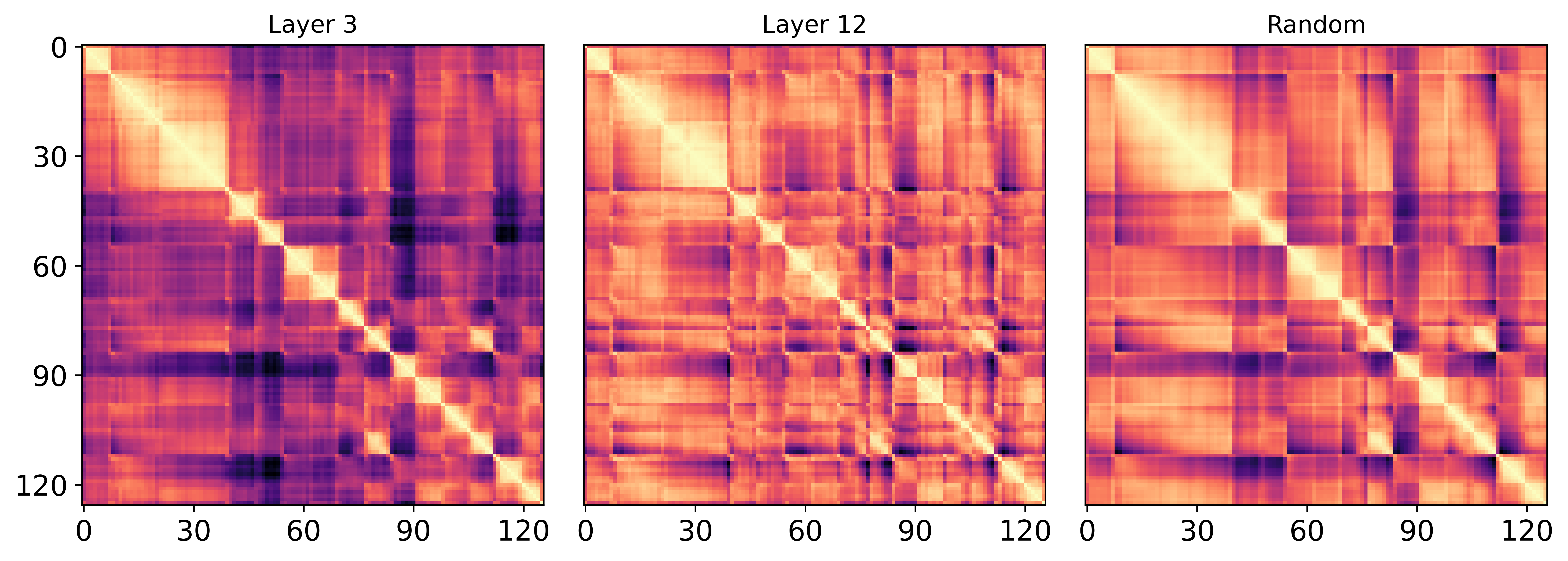}
    \end{subfigure}
    }
    
    \vspace{2mm} 
    \resizebox{\columnwidth}{!}{ 
    \begin{subfigure}{0.5\textwidth}
        \centering
        \includegraphics[width=\textwidth]{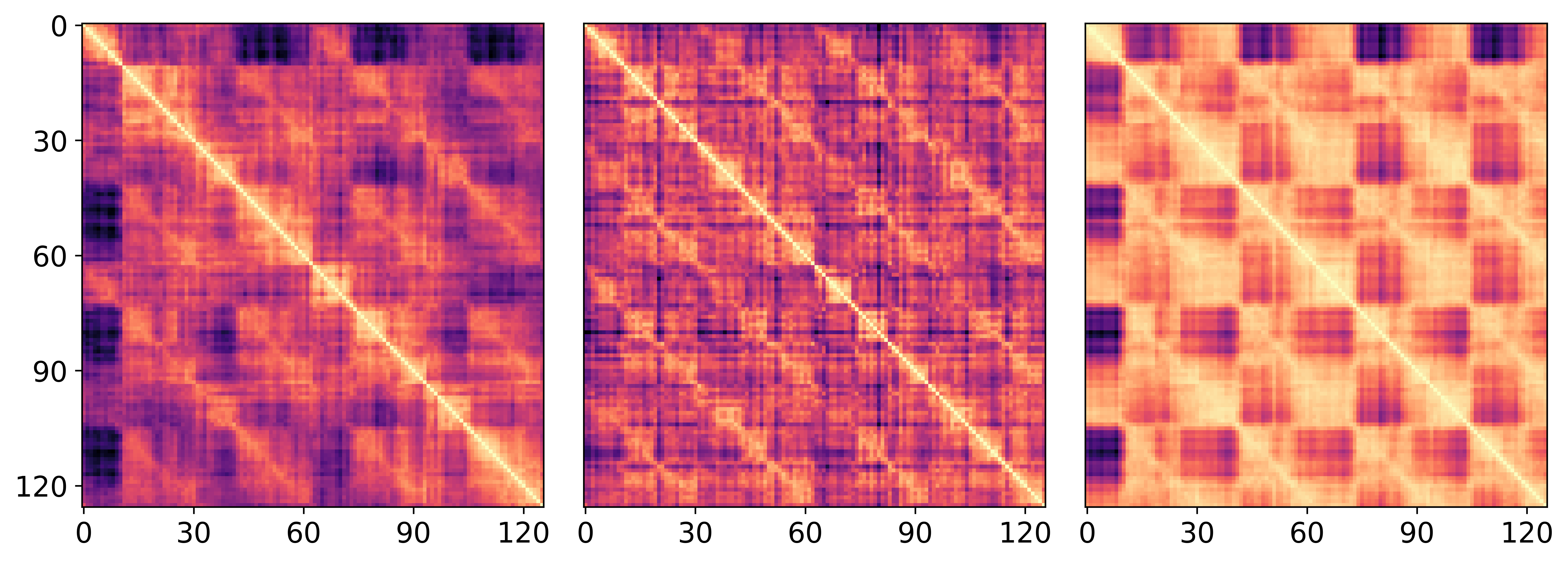}
    \end{subfigure}
    }
    
    \caption{Self-similarity matrices (SSM) of tokens at different layers. Sample 1 (top row) is a melody-dominant audio sample. Sample 2 (bottom row) is a percussion-only sample. From left to right, layer 3, layer 12 and output of a randomly  initializated \melt. We observe clearer block pattern on the SSM of layer 3 (top left) which signifies harmonic properties are captured in the tokens, and clearer subdiagonals on the SSM of layer 12 (bottom middle) which signifies rhythmic properties are more dominant.}
    \label{fig:ssl}
\end{figure}

We show $\boldsymbol{S}_{3}$, $\boldsymbol{S}_{{12}}$, $\boldsymbol{S}_{r}$ on two audio samples of 4 seconds in Figure \ref{fig:ssl} ($\boldsymbol{S}_{{6}}$ and $\boldsymbol{S}_{{9}}$ are omitted due to space limit). Sample 1 (top row) is a monophonic song sample from RWC Pop dataset where a clear melody line is presented. Sample 2 (bottom row) is a sample from ballroom where clear beats are shown by percussive instruments. We observe several properties of tokens from different layers from Figure \ref{fig:ssl}:

\textbf{Randomly initialized model contains harmonic information.} $\boldsymbol{S}_{r}$ for sample 1 contains block structures that correspond to note events, suggesting the model captures harmonic features early on. For sample 2, it fails to capture rhythmic events, which should manifest as evenly spaced subdiagonal structures. Notably, we observe that $\boldsymbol{S}_{r}$ closely resembles the SSM of the model's input, the mel-spectrograms. This observation suggests that more harmonic information is present than rhythmic information in $\boldsymbol{S}_{r}$. The reason might be that the \melt model incorporates skip connections between transformer blocks, which causes the output of a randomly initialized model to closely mirror the mel-spectrogram.

\textbf{Different layers encode different information.} $\boldsymbol{S}_{3}$ exhibits clearer block-like structures than $\boldsymbol{S}_{{12}}$ or $\boldsymbol{S}_{r}$ for sample 1, suggesting that similar harmonic frames are embedded by similar representations. In contrast, $\boldsymbol{S}_{{12}}$, which corresponds to a deeper layer, reveals clearer subdiagonal structures than others for sample 2. These subdiagonals are characteristic of rhythmic patterns that shows the regularity of beats. Shifting from harmonic to rhythmic features could reflect the hierarchical nature of the model, where harmonic features like pitch are learned in shallow layers, while higher-level abstractions and rhythmic features, are learned in the deeper layers. 

\subsection{Downstream results on stacked tokens}
\label{sec:quan-features}

We aim to investigate \textbf{1)} which properties emerge aside from those inherited from the mel-spectrograms; \textbf{2)} if tokens from intermediate transformer blocks are beneficial for downstream tasks.

To achieve this, we use the same downstream training schemes and datasets described in Section \ref{sec:downstream}. Specifically, we evaluate $\boldsymbol{z}_r$ and explore the effect of stacking intermediate tokens [$\boldsymbol{z}_3$, $\boldsymbol{z}_6$, $\boldsymbol{z}_9$, $\boldsymbol{z}_{12}$] to form representations of dimension $192 \times 4 = 768$ for downstream tasks.  As observed in literature, stacking tokens from different layers help with certain tasks, since redundancy of information exist at the same layer\cite{9140343, simoulin-crabbe-2021-many, clark2019what}. We believe using weighted sum of all intermediate layers could further boost performance, we leave that to future work.

Table \ref{tab:random-results} shows that the model at random initialization performs slightly worse than our trained \melt model on chord estimation, much worse on key estimation, but still has a reasonable performance. Harmonic information is by design embedded in mel-spectrograms and is transmitted through the skip connections. For music tagging and beat tracking, there is a significant performance gap between the randomly initialized and trained \melt. This result is expected, as mel-spectrograms are not ideal representations of high-level musical concepts or rhythmic structures.
Moreover, for trained \melt, we observe stacking tokens together significantly improves performance in chord estimation, slightly less in key estimation, but still to a good extent. This indicates that shallow layers contribute to both global and local harmonic tasks. In contrast, performance on music tagging and beat tracking remains similar, suggesting the features captured in shallower layers focus less on rhythmic structures and higher-level musical concepts.

The differences in features learned at various layers highlight that applying a contrastive pretext only to the class token can lead to emergent properties in sequence tokens at different levels.
\begin{table}[t]
    \centering
    \small
    \begin{tabular}{cccccc}
    \toprule
               & \multicolumn{2}{c}{\textsc{Tagging}} & \textsc{Key}               & \textsc{Beat} 
           &  \textsc{Chord}           \\ 
               & \textsc{ROC}            & \textsc{MAP}            & \textsc{w. acc} & \textsc{F-score}        & \textsc{ACC}        \\ \midrule
    \textsc{Random}     & .273          & .807          & .487             & .463          & .290          \\
    \textsc{Train\textsubscript{Last}} & .417          & .896 & .622             & .723 & .319          \\
    \textsc{Train\textsubscript{Stack}}      & \textbf{.437} & \textbf{.902} & \textbf{.639}    & \textbf{.728}         & \textbf{.422} \\ \bottomrule
    \end{tabular}
    \caption{Downstream performance by using randomly initialized \melt, the last layer of a trained \melt (Section \ref{sec:downstream}), and a stack tokens of 4 different layers (Section \ref{sec:quan-features}). The testing datasets and metrics are identical to Section \ref{sec:downstream}.}
    \label{tab:random-results}
\end{table}

\section{Conclusion}

In this paper, we show the ability of a general-purpose contrastive pretext task paired with a transformer to learn local musical representations. Applying NT-Xent loss only to the class token in a lightweight \melt\ surprisingly enables sequence tokens to handle local tasks while contributing to global ones. Despite the class token’s time-invariance, weight sharing and attention mechanisms allow temporal musical representations to emerge.

By analyzing attention maps, we observe that onset events can be deduced.
Self-similarity matrics show different layer tokens capture distinct musical dimensions.
Stacking intermediate tokens improves performance on harmonic tasks, highlighting the importance of shallow-layer representations for downstream tasks.

We provide exploratory insights into the emergent properties of a transformer trained contrastively.
Future work could further study emergent properties in all layers and whether similar emergent properties exist in supervised pretraining. Additionally, leveraging these properties in contrastive pretraining could lead to more efficient pretraining strategies.

\bibliography{ISMIRtemplate}

\begin{thebibliography}{10}
\providecommand{\url}[1]{#1}
\csname url@samestyle\endcsname
\providecommand{\newblock}{\relax}
\providecommand{\bibinfo}[2]{#2}
\providecommand{\BIBentrySTDinterwordspacing}{\spaceskip=0pt\relax}
\providecommand{\BIBentryALTinterwordstretchfactor}{4}
\providecommand{\BIBentryALTinterwordspacing}{\spaceskip=\fontdimen2\font plus
\BIBentryALTinterwordstretchfactor\fontdimen3\font minus \fontdimen4\font\relax}
\providecommand{\BIBforeignlanguage}[2]{{%
\expandafter\ifx\csname l@#1\endcsname\relax
\typeout{** WARNING: IEEEtran.bst: No hyphenation pattern has been}%
\typeout{** loaded for the language `#1'. Using the pattern for}%
\typeout{** the default language instead.}%
\else
\language=\csname l@#1\endcsname
\fi
#2}}
\providecommand{\BIBdecl}{\relax}
\BIBdecl

\bibitem{lostanlen2017convolutional}
V.~Lostanlen, ``Convolutional operators in the time-frequency domain,'' Ph.D. dissertation, École normale supérieure, 2017.

\bibitem{PESTO}
A.~Riou, S.~Lattner, G.~Hadjeres, and G.~Peeters, ``Pesto: Pitch estimation with self-supervised transposition-equivariant objective,'' in \emph{Proc. of theInternational Society for Music Information Retrieval Conference (ISMIR)}, 2023.

\bibitem{quinton2022equivariant}
E.~Quinton, ``Equivariant self-supervision for musical tempo estimation,'' in \emph{Proc. of the International Society for Music Information Retrieval Conference (ISMIR)}, 2022.

\bibitem{kong2024stone}
Y.~Kong, V.~Lostanlen, G.~Meseguer-Brocal, S.~Wong, M.~Lagrange, and R.~Hennequin, ``{STONE}: Self-supervised tonality estimator,'' \emph{Proc. of the International Society for Music Information Retrieval Conference (ISMIR)}, 2024.

\bibitem{kongskey}
Y.~Kong, G.~Meseguer-Brocal, V.~Lostanlen, M.~Lagrange, and R.~Hennequin, ``S-key: Self-supervised learning of major and minor keys from audio,'' in \emph{ICASSP 2025 - IEEE International Conference on Acoustics, Speech and Signal Processing (ICASSP)}, 2025, pp. 1--5.

\bibitem{meseguer2020data}
G.~Meseguer-Brocal, R.~Bittner, S.~Durand, and B.~Brost, ``Data cleansing with contrastive learning for vocal note event annotations,'' in \emph{Proceedings of the 21st International Society for Music Information Retrieval Conference}, 2020.

\bibitem{ma2024foundation}
Y.~Ma, A.~{\O}land, A.~Ragni, B.~M. Del~Sette, C.~Saitis, C.~Donahue, C.~Lin, C.~Plachouras, E.~Benetos, E.~Shatri \emph{et~al.}, ``Foundation models for music: A survey,'' \emph{arXiv preprint arXiv:2408.14340}, 2024.

\bibitem{spijkervet2021contrastive}
J.~Spijkervet and J.~A. Burgoyne, ``Contrastive learning of musical representations,'' in \emph{Proc. of the International Society for Music Information Retrieval Conference (ISMIR)}, 2021.

\bibitem{mccallum2022supervised}
M.~C. McCallum, F.~Korzeniowski, S.~Oramas, F.~Gouyon, and A.~F. Ehmann, ``Supervised and unsupervised learning of audio representations for music understanding,'' 2022.

\bibitem{simclr}
T.~Chen, S.~Kornblith, M.~Norouzi, and G.~E. Hinton, ``A simple framework for contrastive learning of visual representations,'' \emph{CoRR}, 2020.

\bibitem{chen2021exploring}
X.~Chen and K.~He, ``Exploring simple siamese representation learning,'' in \emph{Proc. of the IEEE/CVF Conference on Computer Vision and Pattern Recognition (CVPR)}, 2021.

\bibitem{zhao2022s3t}
H.~Zhao, C.~Zhang, B.~Zhu, Z.~Ma, and K.~Zhang, ``S3t: Self-supervised pre-training with swin transformer for music classification,'' in \emph{ICASSP 2022-2022 IEEE International Conference on Acoustics, Speech and Signal Processing (ICASSP)}.\hskip 1em plus 0.5em minus 0.4em\relax IEEE, 2022, pp. 606--610.

\bibitem{garoufis2023multisource}
C.~Garoufis, A.~Zlatintsi, and P.~Maragos, ``Multi-source contrastive learning from musical audio,'' in \emph{Proc. of the Sound and Music Computing Conference (SMC)}, May 2023.

\bibitem{mccallum2024effect}
M.~C. McCallum, M.~E. Davies, F.~Henkel, J.~Kim, and S.~E. Sandberg, ``On the effect of data-augmentation on local embedding properties in the contrastive learning of music audio representations,'' in \emph{ICASSP 2024-2024 IEEE International Conference on Acoustics, Speech and Signal Processing (ICASSP)}.\hskip 1em plus 0.5em minus 0.4em\relax IEEE, 2024.

\bibitem{choi2022towards}
J.~Choi, S.~Jang, H.~Cho \emph{et~al.}, ``Towards proper contrastive self-supervised learning strategies for music audio representation,'' in \emph{2022 IEEE International Conference on Multimedia and Expo (ICME)}.\hskip 1em plus 0.5em minus 0.4em\relax IEEE, 2022, pp. 1--6.

\bibitem{dhariwal2020jukebox}
P.~Dhariwal, H.~Jun, C.~Payne, J.~W. Kim, A.~Radford, and I.~Sutskever, ``Jukebox: A generative model for music,'' \emph{arXiv preprint arXiv:2005.00341}, 2020.

\bibitem{pasini2024music2latent}
M.~Pasini, S.~Lattner, and G.~Fazekas, ``Music2latent: Consistency autoencoders for latent audio compression,'' \emph{Proc. of the International Society for Music Information Retrieval Conference (ISMIR)}, 2024.

\bibitem{li2023mert}
Y.~Li, R.~Yuan, G.~Zhang, Y.~Ma, X.~Chen, H.~Yin, C.~Lin, A.~Ragni, E.~Benetos, N.~Gyenge, R.~Dannenberg, R.~Liu, W.~Chen, G.~Xia, Y.~Shi, W.~Huang, Y.~Guo, and J.~Fu, ``Mert: Acoustic music understanding model with large-scale self-supervised training,'' in \emph{Proc. of the International Conference on Learning representations (ICLR)}, 2023.

\bibitem{hsu2021hubert}
W.-N. Hsu, B.~Bolte, Y.-S. Chuang \emph{et~al.}, ``Hubert: Self-supervised speech representation learning by masked prediction of hidden units,'' \emph{IEEE/ACM Transactions on Audio, Speech, and Language Processing}, vol.~29, pp. 3451--3460, 2021.

\bibitem{niizumi2024m2dx}
D.~Niizumi, D.~Takeuchi, Y.~Ohishi, N.~Harada, and K.~Kashino, ``{Masked Modeling Duo: Towards a Universal Audio Pre-training Framework},'' \emph{IEEE/ACM Trans. Audio, Speech, Language Process.}, vol.~32, pp. 2391--2406, 2024.

\bibitem{won2024foundation}
M.~Won, Y.-N. Hung, and D.~Le, ``A foundation model for music informatics,'' in \emph{Proc. of the IEEE International Conference on Acoustics, Speech and Signal Processing (ICASSP)}.\hskip 1em plus 0.5em minus 0.4em\relax IEEE, 2024.

\bibitem{wang2022towards}
L.~Wang, P.~Luc, Y.~Wu, A.~Recasens, L.~Smaira, A.~Brock, A.~Jaegle, J.-B. Alayrac, S.~Dieleman, J.~Carreira \emph{et~al.}, ``Towards learning universal audio representations,'' in \emph{ICASSP 2022-2022 IEEE International Conference on Acoustics, Speech and Signal Processing (ICASSP)}.\hskip 1em plus 0.5em minus 0.4em\relax IEEE, 2022, pp. 4593--4597.

\bibitem{koutini2022efficient}
K.~Koutini, J.~Schl{\"u}ter, H.~Eghbal-zadeh, and G.~Widmer, ``Efficient training of audio transformers with patchout,'' in \emph{Proc. Interspeech 2022}, 2022, pp. 2753--2757.

\bibitem{gong21b_interspeech}
Y.~Gong, Y.-A. Chung, and J.~Glass, ``{AST: Audio Spectrogram Transformer},'' in \emph{Proc. of Interspeech}, 2021, pp. 571--575.

\bibitem{manco2022contrastive}
I.~Manco, E.~Benetos, E.~Quinton, and G.~Fazekas, ``Contrastive audio-language learning for music,'' in \emph{Proc. of the International Society for Music Information Retrieval Conference (ISMIR)}, 2022.

\bibitem{huang2022mulan}
Q.~Huang, A.~Jansen, J.~Lee, R.~Ganti, J.~Y. Li, and D.~P. Ellis, ``{MuLan}: A joint embedding of music audio and natural language,'' \emph{Proc. of the International Society for Music Information Retrieval Conference (ISMIR)}, 2022.

\bibitem{dosovitskiy2021an}
A.~Dosovitskiy, L.~Beyer, A.~Kolesnikov, D.~Weissenborn, X.~Zhai, T.~Unterthiner, M.~Dehghani, M.~Minderer, G.~Heigold, S.~Gelly, J.~Uszkoreit, and N.~Houlsby, ``An image is worth 16x16 words: Transformers for image recognition at scale,'' in \emph{International Conference on Learning Representations}, 2021.

\bibitem{caron2021emerging}
M.~Caron, H.~Touvron, I.~Misra, H.~J{\'e}gou, J.~Mairal, P.~Bojanowski, and A.~Joulin, ``Emerging properties in self-supervised vision transformers,'' in \emph{Proc. of the IEEE/CVF international conference on computer vision}, 2021, pp. 9650--9660.

\bibitem{oquab2023dinov2}
M.~Oquab, T.~Darcet, T.~Moutakanni, H.~V. Vo, M.~Szafraniec, V.~Khalidov, P.~Fernandez, D.~Haziza, F.~Massa, A.~El-Nouby, R.~Howes, P.-Y. Huang, H.~Xu, V.~Sharma, S.-W. Li, W.~Galuba, M.~Rabbat, M.~Assran, N.~Ballas, G.~Synnaeve, I.~Misra, H.~Jegou, J.~Mairal, P.~Labatut, A.~Joulin, and P.~Bojanowski, ``Dinov2: Learning robust visual features without supervision,'' 2023.

\bibitem{meseguer2024experimental}
G.~Meseguer-Brocal, D.~Desblancs, and R.~Hennequin, ``An experimental comparison of multi-view self-supervised methods for music tagging,'' in \emph{Proc. of the IEEE International Conference on Acoustics, Speech and Signal Processing (ICASSP)}.\hskip 1em plus 0.5em minus 0.4em\relax IEEE, 2024.

\bibitem{law2009magnatagatune}
E.~Law, K.~West, M.~I. Mandel, M.~Bay, and J.~S. Downie, ``Evaluation of algorithms using games: The case of music tagging,'' in \emph{Proc. of the International Society for Music Information Retrieval Conference (ISMIR)}, 2009, pp. 213--218.

\bibitem{lee2017sample}
J.~Lee, J.~Park, K.~L. Kim, and J.~Nam, ``Sample-level deep convolutional neural networks for music auto-tagging using raw waveforms,'' 2017.

\bibitem{wong2023fmak}
S.~Wong and G.~Hernandez, ``Fmak: A dataset of key and mode annotations for the free music archive--extended abstract,'' in \emph{Proc. of the International Society for Music Information Retrieval Late-Breaking/Demo Session (ISMIR-LBD)}, 2023.

\bibitem{defferrard2017fma}
M.~Defferrard, K.~Benzi, P.~Vandergheynst, and X.~Bresson, ``Fma: A dataset for music analysis,'' \emph{Proc. of theInternational Society for Music Information Retrieval Conference (ISMIR)}, 2017.

\bibitem{knees2022giansteps}
P.~Knees, A.~Faraldo, P.~Herrera, R.~Vogl, S.~B\"ock, F.~H\"orschl\"ager, and M.~Le~Goff, ``Two datasets for tempo estimation and key detection in electronic dance music annotated from user corrections,'' in \emph{Proc. of the International Society for Music Information Retrieval Conference (ISMIR)}, 2015.

\bibitem{raffel2014mir_eval}
C.~Raffel, B.~McFee, E.~J. Humphrey, J.~Salamon, O.~Nieto, D.~Liang, D.~P. Ellis, and C.~C. Raffel, ``Mir\_eval: A transparent implementation of common mir metrics.'' in \emph{Proc. of the International Society for Music Information Retrieval Conference (ISMIR)}, vol.~10, 2014, p. 2014.

\bibitem{gouyon2004ballroom}
F.~Gouyon and S.~Dixon, ``A review of rhythm description systems,'' in \emph{Proc. of the International Society for Music Information Retrieval Conference (ISMIR)}, 2004.

\bibitem{marchand2015gtzanrhythm}
U.~Marchand, Q.~Fresnel, and G.~Peeters, ``Gtzan-rhythm: Extending the gtzan test-set with beat, downbeat and swing annotations,'' in \emph{Proc. of the International Conference on Music Information Retrieval Late-breaking/Demo (ISMIR-LBD)}, 2015.

\bibitem{krebs2015efficient}
F.~Krebs, S.~B{\"o}ck, and G.~Widmer, ``An efficient state-space model for joint tempo and meter tracking.'' in \emph{Proc. of the International Society for Music Information Retrieval Conference (ISMIR)}, 2015, pp. 72--78.

\bibitem{weiss2021schubert}
C.~Wei{\ss}, F.~Zalkow, V.~Arifi-M{\"u}ller, M.~M{\"u}ller, H.~V. Koops, A.~Volk, and H.~G. Grohganz, ``Schubert winterreise dataset: A multimodal scenario for music analysis,'' \emph{Journal on Computing and Cultural Heritage (JOCCH)}, vol.~14, no.~2, pp. 1--18, 2021.

\bibitem{goto2002rwc}
M.~Goto and H.Hashiguchi, ``Rwc music database: Popular, classical, and jazz music databases,'' \emph{Proc. of the International Conference on Music Information Retrieval Conference (ISMIR)}, 2002.

\bibitem{vaswani2017attention}
A.~Vaswani, N.~Shazeer, N.~Parmar, J.~Uszkoreit, L.~Jones, A.~N. Gomez, {\L}.~Kaiser, and I.~Polosukhin, ``Attention is all you need,'' \emph{Advances in neural information processing systems}, vol.~30, 2017.

\bibitem{9140343}
B.~Wang and C.-C.~J. Kuo, ``Sbert-wk: A sentence embedding method by dissecting bert-based word models,'' \emph{IEEE/ACM Transactions on Audio, Speech, and Language Processing}, vol.~28, pp. 2146--2157, 2020.

\bibitem{emiya2009multipitch}
V.~Emiya, R.~Badeau, and B.~David, ``Multipitch estimation of piano sounds using a new probabilistic spectral smoothness principle,'' \emph{IEEE Transactions on Audio, Speech, and Language Processing}, vol.~18, no.~6, pp. 1643--1654, 2009.

\bibitem{2020SciPy-NMeth}
P.~Virtanen, R.~Gommers, T.~E. Oliphant, M.~Haberland, T.~Reddy, D.~Cournapeau, E.~Burovski, P.~Peterson, W.~Weckesser, J.~Bright, S.~J. {van der Walt}, M.~Brett, J.~Wilson, K.~J. Millman, N.~Mayorov, A.~R.~J. Nelson, E.~Jones, R.~Kern, E.~Larson, C.~J. Carey, {\.I}.~Polat, Y.~Feng, E.~W. Moore, J.~{VanderPlas}, D.~Laxalde, J.~Perktold, R.~Cimrman, I.~Henriksen, E.~A. Quintero, C.~R. Harris, A.~M. Archibald, A.~H. Ribeiro, F.~Pedregosa, P.~{van Mulbregt}, and {SciPy 1.0 Contributors}, ``{{SciPy} 1.0: Fundamental Algorithms for Scientific Computing in Python},'' \emph{Nature Methods}, vol.~17, pp. 261--272, 2020.

\bibitem{simoulin-crabbe-2021-many}
A.~Simoulin and B.~Crabb{\'e}, ``How many layers and why? {A}n analysis of the model depth in transformers,'' in \emph{Proc. of the 59th Annual Meeting of the Association for Computational Linguistics and the 11th International Joint Conference on Natural Language Processing: Student Research Workshop}.\hskip 1em plus 0.5em minus 0.4em\relax Association for Computational Linguistics, Aug. 2021, pp. 221--228.

\bibitem{clark2019what}
K.~Clark, U.~Khandelwal, O.~Levy, and C.~D. Manning, ``What does bert look at? an analysis of bert's attention,'' in \emph{BlackBoxNLP@ACL}, 2019.

\end{thebibliography}
\end{document}